\documentclass{jfm-arxiv}

\usepackage{graphicx,color,amsmath,amssymb}
\usepackage{natbib}

\newcommand{\RomanNumeralCaps}[1]
\linenumbers

\usepackage{tabularx}
\usepackage{overpic}
\usepackage{indentfirst}
\usepackage{relsize}

\usepackage{hyperref}
\usepackage[hyphenbreaks]{breakurl}
\hypersetup{colorlinks=true,
			urlcolor=black,
			linkcolor=black,
			citecolor=blue,
			bookmarksdepth=paragraph,
			pdftitle={Lifetimes of metastable windy states in two-dimensional Rayleigh--B\'enard convection with stress-free boundaries (arXiv version)},
			pdfauthor={Qi Wang, David Goluskin, Detlef Lohse}}

\usepackage{cleveref}
\crefname{section}{section}{sections}
\crefname{subsection}{subsection}{subsections}
\crefname{figure}{figure}{figures}
\crefname{table}{table}{tables}
\crefname{equation}{}{}
\Crefname{section}{Section}{Sections}
\Crefname{subsection}{Subsection}{Subsections}
\Crefname{figure}{Figure}{Figures}
\Crefname{table}{Table}{Tables}

\newcommand{\bu}{\mathbf{u}}

\newcommand\solidrule[1][15pt]{\rule[0.5ex]{#1}{1pt}}
\newcommand\dashedrule{\mbox{%
		\solidrule[3pt]\hspace{3pt}\solidrule[3pt]\hspace{3pt}\solidrule[3pt]}}

\title{Lifetimes of metastable windy states in two-dimensional Rayleigh--B\'enard convection with stress-free boundaries}

\author{Qi Wang\aff{1}
\corresp{\email{wangq33@sustech.edu.cn, goluskin@uvic.ca,\\d.lohse@utwente.nl}},
David Goluskin\aff{2}$\dag$
\and Detlef Lohse\aff{3,4}$\dag$
}

\affiliation{\aff{1}Department of Mechanics and Aerospace Engineering, Southern University of Science and Technology, 518055 Shenzhen, China
\aff{2}Department of Mathematics and Statistics, University of Victoria, Victoria, BC, V8P 5C2, Canada
\aff{3}Physics of Fluids Group and Max Planck Center for Complex Fluid Dynamics, J.\ M.\ Burgers Centre for Fluid Dynamics, University of Twente, P.O. Box 217, 7500AE Enschede, The Netherlands
\aff{4}Max Planck Institute for Dynamics and Self-Organization, 37077 G\"ottingen, Germany}

\begin{document}
\maketitle

\begin{abstract}

\noindent Two-dimensional horizontally periodic Rayleigh--B\'enard convection between stress-free boundaries displays two distinct types of states, depending on the initial conditions. Roll states are composed of pairs of counter-rotating convection rolls. Windy states are dominated by strong horizontal wind (also called zonal flow) that is vertically sheared, precludes convection rolls, and suppresses heat transport. Windy states occur only when the Rayleigh number $Ra$ is sufficiently above the onset of convection. At intermediate $Ra$ values, windy states can be induced by suitable initial conditions, but they undergo a transition to roll states after finite lifetimes. At larger $Ra$ values, where windy states have been observed for the full duration of simulations, it is unknown whether they represent chaotic attractors or only metastable states that would eventually undergo a transition to roll states. We study this question using direct numerical simulations of a fluid with a Prandtl number of 10 in a layer whose horizontal period is 8 times its height. At each of seven $Ra$ values between $9\times10^6$ and $2.25\times10^7$ we have carried out 200 or more simulations, all from initial conditions leading to windy convection with finite lifetimes. The lifetime statistics at each $Ra$ indicate a memoryless process with survival probability decreasing exponentially in time. The mean lifetimes grow with $Ra$ approximately as $Ra^4$. This analysis provides no $Ra$ value at which windy convection becomes stable; it might remain metastable at larger $Ra$ with extremely long lifetimes.
\end{abstract}

\begin{keywords}
\end{keywords}


\section{Introduction}
\label{sec:intro}

In the Rayleigh--B\'enard convection (RBC) model, buoyancy-driven flow in a fluid layer is sustained by a destabilizing temperature drop from the bottom boundary to the top one. This system has been studied in laboratory experiments for over a century, and in recent decades it has also been the subject of many direct numerical simulations (DNS) in two and three dimensions with various boundary conditions for the velocity and temperature fields \citep{ahlers2009heat,lohse2010small,chilla2012new,xia2013current,shishkina2021}. In the two-dimensional (2D) case, RBC most often forms convection rolls of alternating rotation direction, with a hot plume rising or a cold plume falling between adjacent rolls. A temperature snapshot for one such pair of rolls is shown in \cref{fig:flow}(e) below. Roll states are seen in simulations with all combinations of no-slip or stress-free velocity boundary conditions, and fixed-temperature or fixed-flux thermal boundary conditions, although the boundary conditions affect what width-to-height ratios the rolls can have \citep{wang2020multiple,wang2020zonal}. Roll states are not the only type of RBC found in 2D, however, at least with certain boundary conditions. 

For 2D RBC that is horizontally periodic and subject to stress-free velocity conditions at the top and bottom boundaries, some simulations have displayed a flow state dominated by a horizontal mean wind. The wind's strong vertical shear suppresses heat transport and precludes convection rolls. \Cref{fig:flow}(b) shows an example of such \emph{windy convection}, sheared so that cold plumes move rightward along the top, while hot plumes move leftward. (The sign of the shear is an arbitrary breaking of symmetry.) The general phenomenon of windy convection, also called zonal flow or shearing convection, has been seen in 2D simulations of various convection models at least as early as \citet{thompson1970} -- see references in \citet[page 363]{goluskin2014convectively}. This windy convection has features in common with strong zonal flows that arise in geophysical and astrophysical systems \citep{richards2006zonal,miyagoshi2010zonal,von2015generation,heimpel2005simulation,kaspi2018jupiter} and in tokamak plasmas \citep{diamond2005}. Although such applications have additional important physics, the 2D RBC model may provide insight as an especially simple system in which convection drives strong zonal flow. Systematic parameter studies of windy convection in 2D RBC were carried out by \citet{goluskin2014convectively} and \citet{wang2020zonal}.

The parameters in the equations modelling RBC can be reduced to two dimensionless numbers: the Rayleigh number $Ra$ that is proportional to the imposed temperature drop across the layer, and the Prandtl number $Pr=\nu/\kappa$, where $\kappa$ and $\nu$ are the fluid's thermal diffusivity and kinematic viscosity, respectively. The aspect ratio $\Gamma$ of the 2D domain is the ratio of the horizontal period to the layer height. At $Ra$ just above the finite value where convection sets in, only roll states exist. For the small horizontal period $\Gamma=2$, roll states become unstable as $Ra$ is raised. There is then a narrow $Ra$ range where the flow seems to switch indefinitely between roll states and windy states with either wind direction \citep{winchester2021}, and at larger $Ra$ only windy states can be found \citep{goluskin2014convectively}. However, the spontaneous transitions from rolls to wind were found to be a small-domain effect by \citet{wang2020zonal}, who simulated flows with horizontal periods as large as $\Gamma=128$. When $\Gamma\ge4$, roll states appear stable for all combinations of $Pr\in[1,100]$ and $Ra\in[10^6,10^9]$ at which simulations were performed, never spontaneously undergoing a transition to windy convection. Windy states were also found in these larger domains at sufficiently large $Ra$; some initial conditions lead to roll states and others to windy states. When $Ra$ is barely large enough to find windy states, these states are transient and eventually undergo a transition to roll states.

In this work we study the spontaneous transition from windy states to roll states. Both states could be called 2D turbulence, with the windy state being only metastable, whereas roll states are apparently stable. We fix $(\Gamma,Pr)=(8,10)$ and seven different $Ra$ values. At each $Ra$ we produce an ensemble of at least 200 simulations from slightly different initial conditions. Every simulation begins with windy convection but eventually undergoes a transition to a roll state with a single pair of rolls. Roll states with multiple pairs of rolls do not seem to arise from windy states, although they can develop from different initial conditions \citep{wang2020zonal}.

Many other fluid systems also display metastable turbulence. Particularly well studied is the spatially localized turbulence in parallel shear flows, which decays to the laminar state at transitional values of the Reynolds number $Re$. Laboratory experiments and DNS have shown that localized ``puffs'' in pipe flow and ``spots'' in planar Couette flow and channel flow have survival probabilities that decrease exponentially in time \citep{bottin1998statistical,faisst2004sensitive,shimizu2019exponential}, similar to what we report below for windy convection. The mean lifetime of a puff or spot, averaged over a large number of instances at each $Re$, appears to increase double-exponentially with $Re$ \citep{hof2008repeller,avila2010transient,shi2013scale,gome2020statistical}. This trend alone does not suggest that shear turbulence becomes truly stable at large $Re$, but a puff or spot has another possible fate: it can split in half, leading to two full-size puffs or spots. While the mean decay time increases superexponentially as $Re$ is raised, the mean splitting time decreases superexponentially. The $Re$ value at which splitting time crosses below decay time has been identified as the onset of sustained turbulence in pipe, Couette, and channel flows \citep{avila2011onset,shi2013scale,gome2020statistical,avila2023}. In relation to the RBC model studied here, decay of a puff or spot is akin to decay of a windy state, but splitting has no analogue. Lacking a mechanism like splitting, if the mean lifetime of wind in RBC remains finite as $Ra$ is raised, then windy convection would not become truly stable, although it could be metastable with extremely long lifetimes.

Decay of metastable turbulence has also been studied in various systems beyond parallel shear flows. \citet{rempel2010supertransient} simulated turbulent-to-laminar decay in magnetized Keplerian shear flow, finding that mean lifetimes increase exponentially with the magnetic Reynolds number. \citet{linkmann2015sudden} simulated turbulent-to-simple-flow decay in isotropic turbulence forced by negative damping at large scales. They find that mean lifetimes increase superexponentially with that system's Reynolds number, as with puff or spot decay in shear flows, but there is no analogue of the splitting mechanism. 

Several systems display transitions from one turbulent state to another, as in our present study, rather than decay to a simple state. \citet{gayout2021rare} report transitions between two turbulent states in wind tunnel experiments where fluid interacts with a pendulum, and \citet{de2022bistability} report transitions into and out of a vortex condensate state in simulations of body-forced turbulence that is triply periodic with a small period in one direction. In both studies the direction of the transition depends on the control parameter. \citet{gayout2021rare} suggest that lifetimes for each transition direction depend double-exponentially on the control parameter. On the other hand, \citet{de2022bistability} suggest that lifetimes diverge at a finite critical value of the control parameter, with the critical value and the rate of divergence differing between the two transition directions.

Metastable turbulence can manifest as switching back and forth between turbulent states, as opposed to permanent disappearance of one state. The present RBC configuration can switch between windy states and roll states in small domains \citep{winchester2021}, as mentioned above, but lifetime statistics have not been studied. In RBC with side walls there is no windy state, but switching occurs between different large-scale circulation patterns in a 2D or quasi-2D square \citep{sugiyama2010flow,chen2019emergence} and in a 3D cylinder \citep{brown2006rotations}. \Citet{chen2019emergence} suggest that mean switching times increase as a power of $Ra$. 

The rest of this paper is organized as follows. \Cref{sec:num} describes the governing equations and our method for simulating an ensemble of flows with transient windy convection. Results are presented in \cref{sec:result}, followed by conclusions in \cref{sec:con}. The appendix describes additional computations that verify the robustness of our results.

\section{Simulation methods}
\label{sec:num}

\noindent 
Rayleigh--B\'enard convection can be modeled by the Boussinesq equations \citep{chandrasekhar1981}, in which the fluid's velocity is divergence-free, and buoyancy force in the vertical $z$ direction is created by the fluid's linear thermal expansion coefficient $\alpha$. In terms of a dimensionless velocity field $\bu(x,z,t)$, temperature field $T(x,z,t)$ and pressure field $p(x,z,t)$, the equations are
\begin{align}
\nabla\cdot\mathbf{u} &= 0, \label{eq:inc} \\
\frac{\partial \mathbf{u}}{\partial t} + \mathbf{u}\cdot\nabla\mathbf{u} 
	&= -\nabla p+ (Pr/Ra)^{1/2}\,\nabla^2\mathbf{u}  + T{\mathbf{e}_z}, \label{eq:u}\\
\frac{\partial T}{\partial t} + \mathbf{u}\cdot\nabla T
	&= (PrRa)^{-1/2}\,\nabla^2 T. \label{eq:T}
\end{align}
The Rayleigh number is $Ra=g\alpha h^3\Delta/\kappa\nu$, where $g$ is gravitational acceleration in the $-z$ direction, $h$ is the layer height, and $\Delta$ is the imposed temperature difference between the top and bottom boundaries. Here $\mathbf{e}_z$ is the unit vector in the $z$ direction. We have scaled length by $h$, so the dimensionless 2D domain is $(x,z)\in[0,\Gamma]\times[0,1]$, with the horizontal $x$ direction being periodic. The dimensionless time $t$ has been scaled by the free-fall time $h/U_f$, where $U_f=(g\alpha h \Delta)^{1/2}$. For the boundary conditions, stress-free conditions on the velocity vector $\bu=(u,w)$ are imposed at both boundaries by $w=0$ and $\partial u/\partial z=0$, and the dimensionless temperatures imposed are $T|_{z=0}=1$ and $T|_{z=1}=0$. We simulated \cref{eq:inc}--\cref{eq:T} with these boundary conditions using the second-order staggered finite difference code AFiD, which has been extensively verified elsewhere; see \cite{verzicco1996finite} and \cite{van2015pencil} for details.

We fix $Pr=10$ because this is the value for which \citet{wang2020zonal} carried out a parameter study of $\Gamma$ and $Ra$. Based on this study we fix $\Gamma=8$ to safely avoid the spontaneous roll-to-wind transitions that occur only in small domains. Simulations are carried out at seven different $Ra$ values between $9\times10^{6}$ and $2.25\times10^7$. The minimum $Ra$ is large enough to induce windy convection, at least initially. The maximum $Ra$ leads to windy convection that can have very long lifetimes but eventually undergoes a transition to rolls in all simulations.

We use the following procedure to create an ensemble of initial conditions that all lead at first to windy convection. At each $Ra$ we start a simulation with temperature that is a random perturbation of the conductive profile $T=1-z$ and with horizontal velocity that is sheared as $u=2(z-1/2)$, where we anticipate that developed flows will have order-unity velocities in our free-fall units. After windy convection develops but before it undergoes a transition to a roll state, we arbitrarily choose one snapshot of the flow. Results are not sensitive to the choice of snapshot (cf.\ the appendix). For the snapshot chosen at each $Ra$, every initial condition in the ensemble is generated by perturbing the temperature at all interior grid points with pseudorandom numbers drawn uniformly from the interval $[-A,A]$. Our main results all use perturbation amplitude $A=10^{-4}$. The appendix reports additional tests confirming that our results are not sensitive to increasing the grid resolution or decreasing the perturbation amplitude.

\section{Results}
\label{sec:result}

\begin{figure}
\centering
\includegraphics[width=0.6\textwidth]{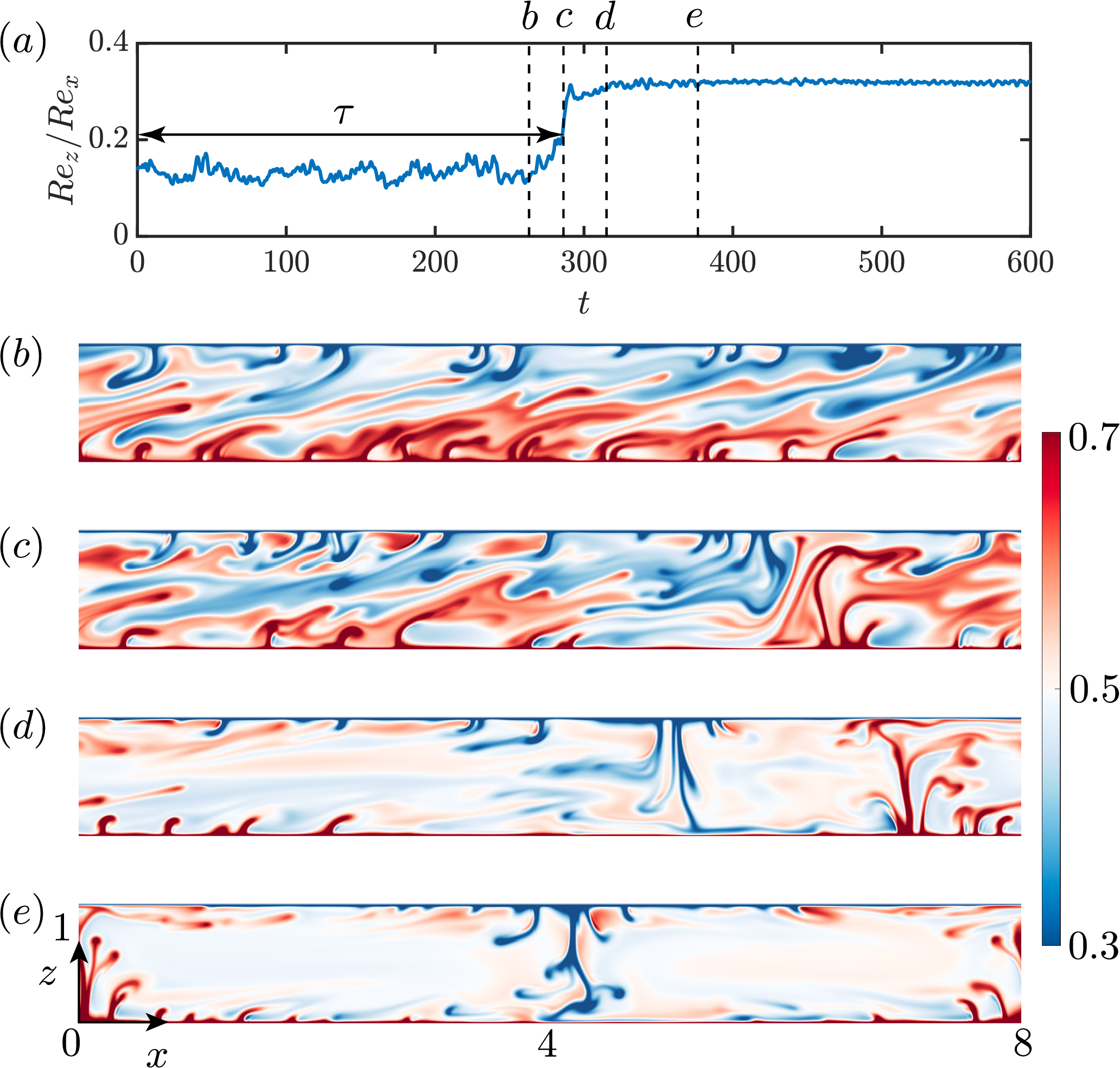}
\caption{\label{fig:flow} ($a$) Time series (in free-fall time units) of the Reynolds number ratio $Re_z/Re_x$ for one simulation with $(\Gamma,Pr,Ra)=(8,10,10^7)$. A transition from the windy state to the roll state occurs around time $\tau\approx286$. ($b$--$e$) Temperature fields at the four instants $t=263.1,286.2,315,376.5$ indicated in panel $a$. The supplementary material includes a movie of the temperature field.}
\end{figure}

\noindent Transitions from the windy state to the roll state occurred in all of our simulations. Each transition was detected as in \citet{wang2020zonal} by using the vertical-to-horizontal ratio of Reynolds numbers, $Re_z/Re_x=\sqrt{\langle w^2\rangle_V/\langle u^2\rangle_V}$, where $\langle\cdot\rangle_V$ denotes a volume average. \Cref{fig:flow} shows one such time series from a simulation where the transition occurred relatively quickly (panel $a$), along with temperature fields before, during and after the transition (panels $b$--$e$). To precisely define the time $\tau$ of a transition, at each $Ra$ we time-averaged $Re_z/Re_x$ in the windy state and in the roll state to find the mean value of each state, then we used the average of these two values as the transition threshold. The first time $Re_z/Re_x$ crosses this threshold defines the lifetime $\tau$ of the simulation. Results are insensitive to the exact threshold; if we instead use the roll-state mean value of $Re_z/Re_x$ as the threshold, the ensemble-averaged lifetime increases by less than 2\% at the smallest $Ra$ (when lifetimes are shortest), and this percentage of increase is smaller at larger $Ra$.

\begin{figure}
\centering
\includegraphics[trim={0 25pt 25pt 6pt},clip,width=0.64\textwidth]{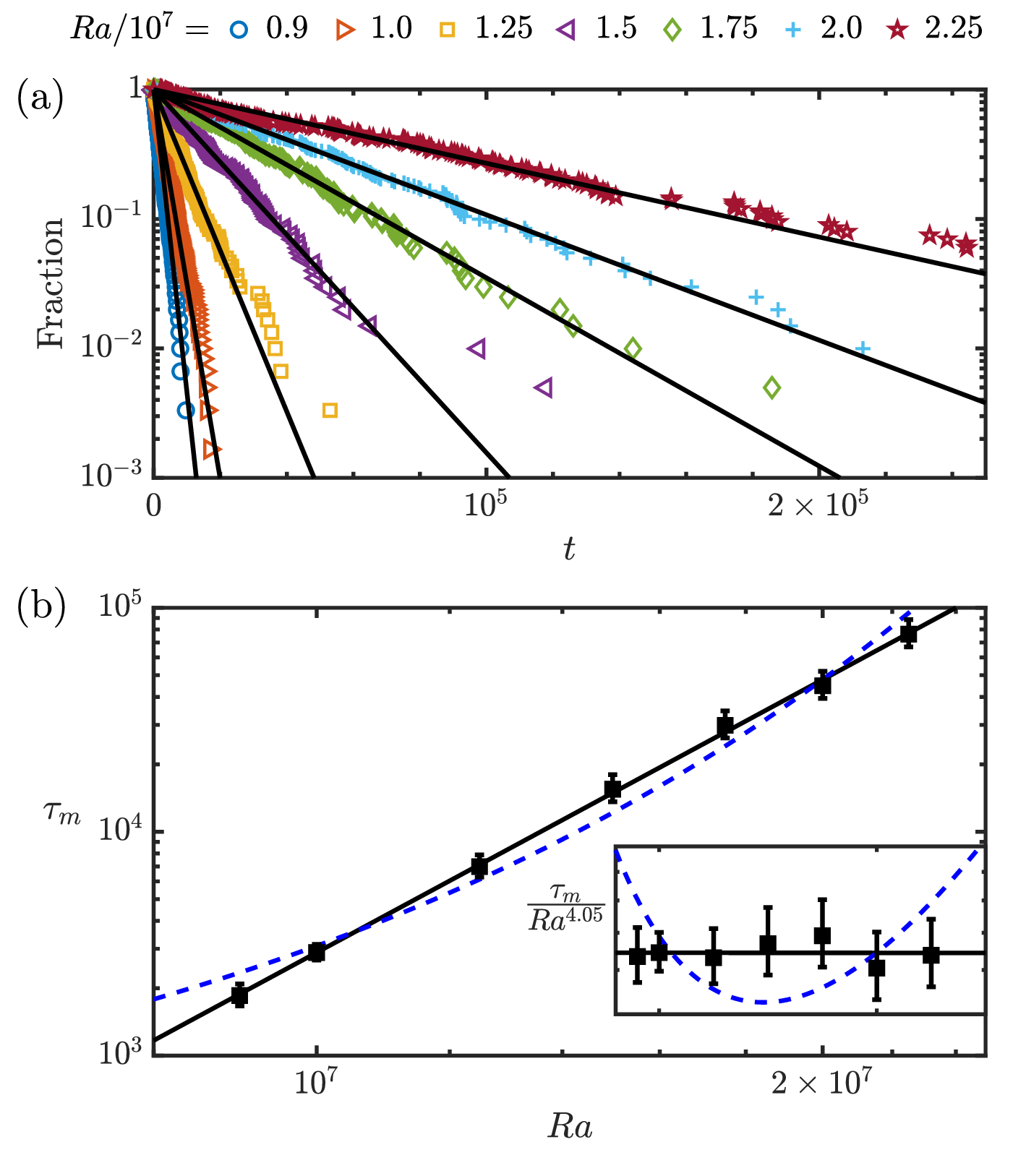}
\caption{\label{fig:lifetimes}(a) Symbols show, for each lifetime $\tau$ of a windy state measured in free-fall times, the fraction of simulations at the same $Ra$ with longer lifetimes. (Some $\tau$ are beyond the plotted timespan.) For the mean lifetime $\tau_m$ at each $Ra$, a solid line shows the survival probability $S(t)=e^{-t/\tau_m}$. (b) Mean lifetimes $\tau_m$ of windy convection ($\mathsmaller{\blacksquare}$) for each $Ra$. Error bars show 95\% confidence intervals (see text). The best-fit power law scaling ($\solidrule$) is $\tau_m\approx c\,Ra^{4.05}$ with $c=1.26\times10^{-25}$. Also shown is the exponential relation obtained by linearly fitting $\log\tau$ to $Ra$ ({\color{blue}$\dashedrule$}), which does not fit the data. The inset shows the same plot with the vertical axis compensated by $Ra^{4.05}$; the range of this axis is $10^{-23}$ to $2\times10^{-23}$.}
\end{figure}

To examine lifetime statistics of windy convection within each ensemble, for every time $\tau$ at which a simulation undergoes a transition, we calculate the fraction of simulations that survive longer than $\tau$. \Cref{fig:lifetimes}(a) shows these fractions plotted versus $\tau$, with a different data series for each $Ra$. Each plotted series is close to a straight line, and the vertical axis scale is logarithmic, so this indicates that the fractions decrease exponentially in time. Such exponential decrease suggests that the wind-to-roll transitions behave as a memoryless random process, as in most other studies of metastable turbulence recalled in the introduction. For a memoryless random process with mean lifetime $\tau_m$, the probability of a chosen ensemble member surviving past time $t$ is exactly $S(t)=e^{-t/\tau_m}$. The straight lines in \cref{fig:lifetimes}(a) show this $S(t)$ for the $\tau_m$ values estimated at the various $Ra$, meaning the lines have slopes of $-1/\tau_m$. Each $\tau_m$ is estimated simply as the average over all lifetimes $\tau$ in an ensemble, which is possible because we have run every simulation until it undergoes a transition. Estimating $\tau_m$ instead from the slope of the data in \cref{fig:lifetimes}(a) gives similar values, as reported in the appendix.

\Cref{fig:lifetimes}(b) shows how the mean lifetimes $\tau_m$ vary with $Ra$. Both axes are logarithmic, so a linear trend corresponds to a power law scaling of $\tau_m$ with $Ra$. The best-fit line gives the scaling $\tau_m\approx c\,Ra^{4.05}$. Error bars plotted for each $\tau_m$ estimate are $\pm1.96\tau_m/\sqrt{N}$, which is the 95\% confidence interval for an $N$-member ensemble from an exponential random process \citep[cf.][]{avila2010transient}. The best-fit scaling exponent of $4.05$ is indistinguishable from 4, according to the 95\% confidence interval $\pm0.14$ that the MATLAB function {\tt confint} estimates from our $\tau_m$ values. \Cref{fig:lifetimes}(b) shows the line of this best-fit scaling law, along with the curve of an exponential fit. The exponential curve clearly does not fit the data well, and any fits with double-exponential or divergent dependence on $Ra$ would be even worse.

\section{Discussion and conclusions}
\label{sec:con}

This first investigation of the lifetime statistics of windy convection in 2D stress-free RBC has been specific to flow with $Pr=10$ and a horizontal period 8 times the layer height. Over the studied range of $9\times10^6\le Ra\le 2.25\times10^7$, the wind-to-roll transition behaves as a memoryless random process, and the ratio of mean wind lifetime to free-fall time scales approximately like $Ra^4$. The corresponding ratio of mean lifetime to diffusive time scales like $Ra^{3.5}$. While $Ra$ varies only by a factor of 2.5 over the range of our simulations, the mean lifetimes vary by a factor of almost 40. This range is large enough to see clearly that the mean lifetimes are better described by power-law scaling with $Ra$ than by exponential or superexponential dependence on $Ra$. Whether this scaling differs at other Prandtl numbers and horizontal periods remains to be studied; both parameters are known to affect the $Ra$ values at which windy states occur in RBC \citep{wang2020zonal} as well as in penetrative convection \citep{Fuentes2021}. Power-law dependence on $Ra$ has been suggested also for mean switching times between different large-scale circulation patterns in RBC \citep{wang2018mechanism,chen2019emergence}.

If power-law scaling of mean lifetimes continues as $Ra\to\infty$, this would mean that windy convection is never truly stable but is metastable with such long lifetimes at large $Ra$ as to be effectively stable. However, the history of transitional shear flow studies suggests caution when extrapolating our findings to asymptotically long lifetimes; for turbulent puffs and slugs, the double-exponential dependence of lifetimes on $Re$ is technically hard to determine, and earlier studies with less data and smaller domains suggested other dependence on $Re$ -- see discussion in \citet{avila2010transient}. Extending our DNS approach to larger $Ra$ would be very expensive; nearly 0.47 million CPU hours were needed to observe transitions in all 200 of our highest-$Ra$ simulations. Somewhat larger $Ra$ could be reached if simulations are truncated at a maximum timespan rather than waiting for every one to undergo a transition; mean lifetime estimates can account for such truncation, as described in \citet{avila2010transient}. At even larger $Ra$ where mean lifetimes are extremely long, one might employ a rare-event algorithm where DNS is performed selectively for cases leading to a transition while pruning others, as has been done for shear flows \citep[e.g.,][]{gome2022}. Another possibility is a laboratory experiment, which is well suited to long time spans, but 2D convection with stress-free top and bottom would be hard to approximate. For this reason and others, it is of great interest to find a laboratory model that displays windy convection.

Regarding a theoretical explanation of how mean lifetimes of windy convection depend on $Ra$, further data may allow an argument that invokes extreme value statistics. For pipe flow, \citet{goldenfeld2010extreme} suggested that puff decay is triggered when the maximum-over-space turbulent intensity drops below some threshold. They note that if this maximum intensity follows a Gumbel distribution, and if the threshold decreases linearly in $Re$ as $Re$ is raised, then the mean time before falling below the threshold would increase double-exponentially with $Re$. For RBC, the wind-to-roll transition mechanism seems the opposite: rolls might be triggered when the maximum-over-space deviation from the mean wind rises \emph{above} some threshold. If this threshold increases linearly in $Ra$, then arguments analogous to \citet{goldenfeld2010extreme} imply that mean lifetime would increase exponentially with $Ra$, not double-exponentially. However, if the threshold increases only logarithmically with $Ra$, then the mean lifetime would increase as a power of $Ra$, as we observe in our data. Further DNS is needed to determine whether such extreme value arguments apply to the present model, as has been largely confirmed for pipe flow by \cite{nemoto2021}. In particular, are wind-to-roll transitions triggered when maximum turbulent intensity rises above a threshold? If so, which extreme value distribution governs this maximum, and how does the threshold vary with $Ra$?

Another direction for simplified modeling of windy convection is as a stochastic predator--prey system. A model of this type has been proposed for pipe flow \citep{shih2016ecological}, where it captures double-exponentially varying mean lifetimes of not only puff decay but also splitting. This model consists of stochastic predator--prey dynamics in a spatially extended domain, with the predator representing azimuthal zonal flow in the pipe and the prey representing turbulent fluctuations. A similar predator--prey analogy has been described for windy convection \citep{goluskin2014convectively}, again with the zonal flow as the predator, but there is no clear analogue to the spatial localization and splitting of puffs in pipe flow. As with the extreme value arguments, closer analysis of various quantities in windy convection may be needed to formulate an insightful stochastic model.

\section*{Acknowledgements}
We thank Roberto Verzicco, Nigel Goldenfeld and Hong-Yan Shih for valuable conversations, and we thank the anonymous referees for helpful comments. Prior to our investigation, D.G.\ discussed preliminary plans with Nigel and Hong-Yan, and Hans Johnston carried out preliminary simulations (unpublished) with a horizontal period twice the layer height. The present simulations were a separate effort, carried out on the national e-infrastructure of SURFsara, a subsidiary of SURF cooperation, the collaborative ICT organization for Dutch education and research. D.G.\ was supported by the Canadian NSERC Discovery Grants Program via award numbers RGPIN-2018-04263, RGPAS-2018-522657, and DGECR-2018-0037. D.L.\ acknowledges funding from the European Research Council (ERC) under the European Union’s Horizon 2020 research and innovation programme (Grant agreement No. 10109442).

\section*{Declaration of interests}

The authors report no conflict of interest.

\section*{Appendix}

\Cref{tab:data} summarizes the main ensembles of simulations on which we have reported above. The full lifetime data sets for these ensembles are given in the supplementary material. In addition to the $\tau_m$ values estimated by averaging all lifetimes in each ensemble, the table reports alternative estimates where $-1/\tau_m$ is the slope of lines fit to the data in \cref{fig:lifetimes}(a). The latter $\tau_m$ values have a best-fit scaling of $Ra^{4.11}$ instead of $Ra^{4.05}$.

\tabcolsep 8pt
\begin{table}
\begin{center}
\begin{tabular}{rccrrrrr}
$Ra$\quad~	& resolution   &     $N$	& $\tau_m$ mean & $ \tau_m$ fit\,
	& $\max\tau$ \\[4pt] 
$9\times10^6$		& $1536\times192$	& 300 & 1854~~ & 1842~ & 9687 \\ 
$10^7$			& $1536\times192$	& 600 & 2887~~ & 2910~ & 16\,850 \\ 
$1.25\times10^7$	& $2048\times256$	& 300 & 6980~~ & 7404~ & 52\,985 \\ 
$1.5\times10^7$	& $2048\times256$	& 200 & 15\,476~ & 16\,479 & 117\,797 \\ 
$1.75\times10^7$	& $2048\times256$	& 200 & 29\,855~ & 30\,142 & 185\,809 \\ 
$2\times10^7$		& $2048\times256$	& 200 &  44\,853~ & 47\,813 & 373\,524 \\ 
$2.25\times10^7$	& $2048\times256$	& 200 &  76\,213~ & 80\,827 & 470\,944 \\ 
\end{tabular}
\caption{\label{tab:data}For each of the main simulation ensembles, columns from left to right give the Rayleigh number, the horizontal and vertical grid resolution, the number of simulations ($N$), the mean lifetime $\tau_m$ estimated by the mean of $\tau$ in the ensemble, $\tau_m$ estimated by fitting a line to data in \cref{fig:lifetimes}(a), and the maximum lifetime in each ensemble. Times are in free-fall units. In all cases, $Pr=10$, the horizontal period is 8 times the layer height, and random perturbations of the initial temperature have amplitude $A=10^{-4}$.}
\end{center}
\end{table}

In the $Ra=10^7$ case, where our main ensemble size of $N=600$ is the largest, we have verified that our grid resolution is adequate, and that results are insensitive to how initial conditions are generated. As reported in \cref{tab:data}, this main ensemble was simulated with a resolution of $1536\times192$ using initial conditions generated from a single snapshot by random temperature perturbations of amplitude $A=10^{-4}$. We performed $N=500$ additional simulations on the finer grid $2048\times256$. The average over these lifetimes gives $\tau_m\approx2806$, which agrees with the main ensemble's value $\tau_m\approx2886$ to within the $\pm8\%$ margin of the 95\% confidence interval. This indicates that the resolution $1536\times192$ is adequate. We performed $N=800$ additional simulations on the coarser grid $1024\times128$. The average over these lifetimes gives $\tau_m\approx2332$. This is 19\% shorter than $\tau_m$ from the main ensemble, which indicates that $1024\times128$ is under-resolved and produces bias towards shorter lifetimes. Finally, at the same resolution as the main ensemble, we performed $N=300$ simulations whose initial temperature perturbations had the smaller amplitude of $A=10^{-5}$. The resulting lifetimes are exponentially distributed, as in all other cases, and their average gives the very similar estimate of $\tau_m\approx2864$.

The same simulations used to confirm that $\tau_m$ is robust to increased resolution and to decreased perturbation amplitude also confirm that $\tau_m$ is robust to the flow snapshot from which initial conditions are generated by random perturbations. This is because the ensembles with medium and high resolutions with $A=10^{-4}$ used different snapshots, and so did the ensemble with medium resolution and $A=10^{-5}$. As reported above, all three snapshots led to similar $\tau_m$ estimates.

\bibliographystyle{jfm}
\bibliography{reference}

\end{document}